# Target-driven merging of Taxonomies

Salvatore Raunich, Erhard Rahm

*University of Leipzig*
*Germany*
`{raunich, rahm}@informatik.uni-leipzig.de`

*Abstract*— The proliferation of ontologies and taxonomies in many domains increasingly demands the integration of multiple such ontologies. The goal of ontology integration is to merge two or more given ontologies in order to provide a unified view on the input ontologies while maintaining all information coming from them. We propose a new taxonomy merging algorithm that, given as input two taxonomies and an equivalence matching between them, can generate an integrated taxonomy in a fully automatic manner. The approach is target-driven, i.e. we merge a source taxonomy into the target taxonomy and preserve the structure of the target ontology as much as possible. We also discuss how to extend the merge algorithm providing auxiliary information, like additional relationships between source and target concepts, in order to semantically improve the final result. The algorithm was implemented in a working prototype and evaluated using synthetic and real-world scenarios.

## I. INTRODUCTION

Ontologies and taxonomies are increasingly used to semantically categorize or annotate information, especially on the web. For example, product catalogs of online shops, comparison portals or web directories categorize products or websites to help users and applications finding relevant information. In life sciences, ontologies are used to describe components and functions of organisms or objects such as genes or proteins. Since many ontologies refer to the same domain and to the same objects, there is a growing need to integrate or merge such related ontologies. The goal is to create a merged ontology providing a unified view on the input ontologies while maintaining all information coming from them.

Despite some previous work, ontology integration is still a challenge, in particular if one wants to perform the integration in a largely automatic way. The related research problem of schema integration has been studied for a long time [4] but most earlier approaches suffered from trying to solve the complex problems of matching and merging in a single approach. More recent work on schema integration builds on the research results on semi-automatic schema matching [15] and separate matching from merging. Hence, several algorithms have been proposed to merge schemas based on a pre-determined match mapping [5], [17], [12], [18], [14]. Despite this simplification, several of these merge approaches are still not fully automatic but depend on manual intervention. Previous approaches on ontology merging [11], [9], [19] are also user-controlled and do not utilize the separation of matching and merging. While user-controlled approaches provide flexibility for determining the merge result, they require the involvement of expensive data integration experts and introduce a substantial manual effort especially for large ontologies.

We therefore propose a new approach for ontology merging which is fully automatic and which utilizes a match mapping between the input ontologies. We propose a target-driven algorithm, i.e. we merge a source taxonomy into the target taxonomy. Such an asymmetric merge is highly relevant in practice and allows us to incrementally extend the target ontology by additional source ontologies. For example, the catalog of a new online shop may be merged into the catalog of a price comparison portal.

*Example 1.1:* Let us consider the example in Figure 1 where the source and target taxonomies represent the product catalogs of two different computer and hardware shops. We suppose that an equivalence matching between concepts is already given, automatically generated by a matching tool or manually designed by an expert user. As we can see, only some source concepts have an equivalent concept in the target. Merging requires the equivalent concepts to be combined. One of the main problems is deciding which remaining concepts should be integrated in the result and what are the best positions of these concepts in the integrated taxonomy. The example shows that the two taxonomies organize hardware products in different ways. The target initially categorizes first by manufacturer and then by product type (laptops, desktops, etc.) while the source taxonomy uses the opposite order. Maintaining both views in the merged taxonomy would introduce semantic overlap and reduce the understandability of the resulting taxonomy. We deal with such situations by giving preference to the target taxonomy. This also allows us to automatically find suitable merge decisions.

The main contribution of this work are new target-driven algorithms to automatically integrate taxonomies. The base algorithm takes as input two taxonomies and an equivalence matching between concepts. We also present an extended algorithm that can utilize additional relationships between the input taxonomies to semantically improve the merge result. The algorithms generate taxonomies that preserve all instances of the input taxonomies as well as the structure of the target taxonomy. In contrast to previous work, we do not necessarily preserve all source concepts but aim at limiting the semantic overlap in the merged taxonomy for improved understandability. This is achieved by utilizing the input mapping and giving preference to the target taxonomy when the same concepts are differently organized in source and target. The algorithms have been implemented in a working prototype and evaluated

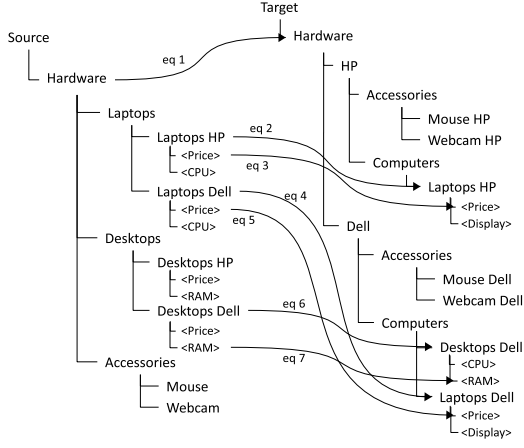

Fig. 1. Running Example

on large real-life ontologies [16].

In the next section, we introduce our ontology model and define the problem. In Section III, we describe the base algorithm in detail and discuss its complexity. Section IV outlines the extended merge algorithm. Section V sketches the generation of mappings between the input ontologies and the merge result that can be used for instance migration. In Section VI we evaluate the algorithms on real-life ontologies. Related work is described in Section VII before we conclude.

## II. MODELS AND PROBLEM DEFINITION

In this section we define data representation models used in the paper. An ontology is a quadruple $O = (C, C_i, I, R)$ where $C$ is a collection of *Classes* or *Concepts*, $C_i \subseteq C$ is the subset of concepts containing instances, $I$ is the set of instances, possibly empty, and $R$ is the set of *relationships* between concepts. A Concept represents a collection of objects with similar properties. Each Concept $C$ has a name (or *label*) and a collection of attributes $A_c$, possibly empty. Several kinds of relationships can be defined, like "is-a" or "subclass", "part-of", "type-of", etc. A relationship $r(a,b) \in R$ is a directed, binary and semantic connection between two concepts $a$ and $b$. It can be explicitly present in the ontology or *implied* by an ontology rule. For example, given two is-a relationships $r(a,b)$ and $r(b,c)$, the relationship $r(a,c)$ is implied since is-a relationships are transitive.

Graphically, as we can see in Figure 1, we represent concepts with a simple label and attributes with a tagged label; we use a nesting notation to represent is-a relationships. For example, in the source ontology, *Laptops Dell* is a concept with attributes *Price* and *CPU*; it is also a subclass of the concept *Laptops*. It is important to note that, in general, a taxonomy is a graph but we use a tree-style representation to simplify the visualization.

In this paper, we will consider only ontologies $O = (C, C_i, I, R)$ where $C_i$ contains only leaf nodes and $R$ contains only "is-a" relationships between concepts. For this reason, in the following, we will use the terms *ontology* and *taxonomy* with the same meaning. Our taxonomies are acyclic but multiple inheritance is supported, i.e. a concept can have multiple parents. In the following we discuss some examples to show how the algorithm can deal with both single and multiple inheritance. Our algorithm can be extended to support instances for inner concepts but we will not provide the details here.

The equivalence mapping between two ontologies $S = (C_s, C_{is}, I_s, R_s)$ and $T = (C_t, C_{it}, I_t, R_t)$ is defined as a set of *correspondences*. We distinguish two different kinds of correspondences: *concept correspondences* and *attribute correspondences*. Given two concepts $s \in C_s$ and $t \in C_t$, we define a concept correspondence $(s,t)$ as an ordered pair of a source concept and a target concept. Similarly, given two attributes $a_s \in A_s$ and $a_t \in A_t$, we define an attribute correspondence $(a_s, a_t)$ as an ordered pair of a source attribute and a target attribute. Figure 1 shows seven example correspondences: $eq_1$, $eq_2$, $eq_4$ and $eq_6$ are concept correspondences, $eq_3$, $eq_5$ and $eq_7$ are attribute correspondences. In the following we will consider only 1:1 equivalence mappings but the algorithm can be extended to support, in general, m:n correspondences. In this paper we do not investigate how these correspondences are generated but we assume that a *correct* and *complete* matching is already given; it can be automatically generated by a matching tool or manually provided by a domain expert user. For the extended algorithm in Section IV we will also consider *is-a* and *inverse-isa* mappings between the input ontologies.

### A. Properties of the merge result

Based on [13], we identified, adapted and extended some properties that the solution of our merge algorithm should satisfy. Called $S = (C_s, C_{is}, I_s, R_s)$ and $T = (C_t, C_{it}, I_t, R_t)$ the input source and target taxonomies and $T' = (C_{t'}, C_{it'}, I_{t'}, R_{t'})$ the merge result, and called $Map_{ST}$, $Map_{ST'}$ and $Map_{TT'}$ the input equivalence mapping between $S$ and $T$ and the generated equivalence mappings between $S$ and $T'$ and $T$ and $T'$ respectively, we have as follows:

**(P1) Target Element Preservation.** Each element (a concept or an attribute) in the input target taxonomy T has a corresponding element in the merge result T'. Formally, each concept $c \in C_t$ corresponds to exactly one concept $c' \in C_{t'}$. This concept correspondence is defined as $(c, c') \in Map_{TT'}$. Similarly, each attribute $a \in A_c$ corresponds to exactly one attribute $a' \in A_{c'}$. This attribute correspondence is defined as $(a, a') \in Map_{TT'}$.

**(P2) Target Relationship Preservation.** Each input target is-a relationship is explicitly in or implied by $T'$. Formally, for each is-a target relationship $r(s,t) \in R_t$, if $(s, s') \in Map_{TT'}$ and $(t, t') \in Map_{TT'}$, then either $r(s', t') \in R_{t'}$ or $r(s', t')$ is implied in $T'$.

**(P3) Information Preservation.** In addition to target maintenance, we require that all instances of both the target and the source ontology must be preserved in the merged taxonomy.

**(P4) Control of semantic overlap.** The merge algorithm should generate an integrated taxonomy with little or no redundancy compared to the input taxonomies. In particular, 1) no instance overlap between concepts should be introduced, i.e. every instance should migrate to exactly one concept in the merge result. Furthermore, 2) we want to avoid or limit multiple paths to leaf nodes introduced by different concept organizations in the input ontologies. Formally, 1) for each instance $i \in I_s \cup I_t$, called $f$ the transformation function that moves $i$ in $I_{t'}$ and defined $i' \in I_{t'}$ as $i' = f(i)$, then such a function $f$ exists and $i'$ is unique in $I_{t'}$; moreover, called $c \in C_s \cup C_t$ and $c' \in C_{t'}$ the concepts containing $i$ and $i'$ respectively, there must exist a correspondence $(c, c') \in Map_{ST'} \cup Map_{TT'}$; 2) for each pair of concepts $c \in C_{it}$ and $c' \in C_{t'}$ such that there exists $(c, c') \in Map_{TT'}$, called $p$ and $p'$ the number of different paths to $c$ and $c'$ respectively in $T$ and $T'$, then $p = p'$.

We remark that if $T$ is a tree-structured taxonomy, i.e. $T$ has no multiple inheritance, then the result $T'$ will remain a tree and no multiple inheritance will be introduced. In the example shown in figure 1, for the leaf concept *Laptops HP* in $T$ we will not take over the different path in $S$ to this concept.

**(P5) Equality preservation.** If two concepts are equal in the equivalence mapping then they are mapped to the same merged concept in the result and vice versa. Formally, if two concepts $s, t \in C_s \cup C_t$ are equal in $Map_{ST}$, then there exists a unique concept $c \in C_{t'}$ such that $(s, c) \in Map_{ST'}$ and $(t, c) \in Map_{TT'}$. If $s$ and $t$ are not equal in $Map_{ST}$, then such a concept $c$ does not exist and $s$ and $t$ correspond to different elements in $T'$.

Finally, we require that the algorithm must terminate and produce a result that is itself a taxonomy (respectively *Termination* and *Closure*) and should be *scalable* and able to provide good performance and acceptable execution times also for large taxonomies with many concepts and is-a relationships.

There is a certain trade-off between requirements (P3) and (P4) for inner concepts in the input taxonomies. Why we could have required that all concepts of both input taxonomies should be preserved for completeness this would often lead to a significant semantic overlap and redundancy in the merged taxonomy reducing its understandability and value. We therefore only require preservation of the target taxonomy and will drop some inner concepts from the source taxonomy that would introduce redundant paths to the leaf nodes in the merged taxonomy.

Based on these properties, we define the ontology merging problem as follows: given as input two taxonomies $S$ and $T$ and an equivalence mapping between them, generate as output a new taxonomy $T'$ satisfying the requirements P1-P5. Furthermore, two mappings between $S$ and $T'$ and between $T$ and $T'$ should be determined specifying where in $T'$ the elements of the input ontologies are mapped.

The ontology merging problem can have many possible solutions and the best one depends on the reference context; usually the evaluation of an integrated ontology is subjective. We aim at an automatic and target-driven merge approach that integrates the first (source) input ontology into the second one and gives preference to the target ontology when equivalent concepts are differently organized in the input ontologies. Our approach generates a default solution in a fully automatic way that may interactively be adapted by users if needed.

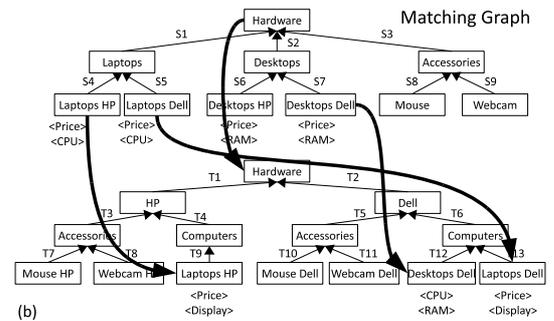

Fig. 2. Matching Graph

III. BASE ALGORITHM

We propose a base algorithm that, given as input two ontologies and an equivalence mapping, produces an integrated ontology that meets the requirements introduced in Section II. For convenience, we split the description of the algorithm in two successive phases: a *preliminary phase* that, starting from the source ontologies and the set of correspondences, generates a so-called integrated concept graph; and a *main phase* that, starting from the integrated concept graph produced before, generates the final result. We focus on generating the merged taxonomy in this section; the generation of the output mappings is explained in Section V.

*A. Preliminary phase*

The preliminary phase, also called *integrated concept graph generation algorithm*, takes as input two ontologies $O_1$ and $O_2$ and an equivalence matching between them, provided as a set of concept correspondences and attribute correspondences, and it generates as output an integrated concept graph $I$. A similar algorithm was introduced in [5] for relational and XML schemas and the preliminary phase of our approach is based on it but with significant differences that we will discuss later in this section. We identify the following steps:

- Step 1.1: Building of the Concept Graphs $G_1$ and $G_2$
- Step 1.2: Building of the Matching Graph $M$
- Step 1.3: Generation of the Integrated Concept Graph $I$

The pseudo-code for the two last steps is described in Algorithms 1 and 2. We omit the pseudo-code for the generation of concept graphs for the input ontologies since it is trivial.

The first step 1.1 transforms the input ontologies $O_1$ and $O_2$ into two different directed concept graphs $G_1$ and $G_2$, respectively, where nodes are concepts and edges are is-a relationships.

In the second step 1.2, a matching graph is produced according to Algorithm 1. Starting from the concept graphs $G_1$ and $G_2$ generated in the previous step, we translate all

**Algorithm 1** $MatchingGraphGen(G_1, G_2, Corr)$

**Input:** two concept graphs $G_1 = (V_1, E_1)$ and $G_2 = (V_2, E_2)$ and a set of equivalence correspondences $Corr$
**Output:** a Matching Graph $M$
1: $M = (V, E) \leftarrow$ empty
2: $V \leftarrow V_1 \cup V_2$
3: **for each** concept correspondence $c = c_1 - c_2$ in $Corr$ **do**
4:   **if** $E$ does not contain an edge between $c_1$ and $c_2$ **then**
5:     add a new edge in $E$ between $c_1$ and $c_2$
6:   **end if**
7: **end for**
8: **return** $M$

---

**Algorithm 2** $ICGGen(O_1, O_2, Corr)$

**Input:** two ontologies $O_1$ and $O_2$ and a set of equivalence correspondences $Corr$
**Output:** an Integrated Concept Graph $I$
1: $I = (V, E) \leftarrow$ empty
2: $G_1 \leftarrow ConceptGraphGen(O_1)$
3: $G_2 \leftarrow ConceptGraphGen(O_2)$
4: $M \leftarrow MatchingGraphGen(G_1, G_2, Corr)$
5: **for each** connected component $CC$ in $M$ **do**
6:   generate an integrated concept $C$
7:   $label(C) \leftarrow genLabel(CC)$
8:   $atts \leftarrow genAttributeList(CC)$
9:   add all attributes in $atts$ to $C$
10:   add $C$ to $V$
11: **end for**
12: **for each** edge $e_1$ in $G_1$ **do**
13:   generate a *s-edge* in $I$
14: **end for**
15: **for each** edge $e_2$ in $G_2$ **do**
16:   generate a *t-edge* in $I$
17: **end for**
18: **return** $I$

---

the input correspondences into *matching edges*. In particular, for each concept correspondence we create a matching edge in the matching graph and for each attribute correspondence we create a matching edge between related concepts, but only if a similar edge has not already been generated. Figure 2 shows the matching graph for our running example of Fig.1; correspondences drawn with a bold line represent matching edges.

The result of the last step 1.3 is an integrated concept graph $I = (V, E)$ and it is generated according to Algorithm 2. First we identify all connected components in the matching graph $M$ with respect to matching edges and for each connected component in $M$ we generate an integrated concept $C$ with a label $l$ depending on merged source concepts. For each integrated concept we save the collections of corresponding source and target concepts. It is important to note that, in the case of 1:1 mappings, a connected component can contain only one concept, for example $c$, if $c$ is not involved in any correspondence or two concepts, for example $s$ and $t$ (with $s \in S$ and $t \in T$), if there exists an equivalence correspondence $(s, t) \in Map_{ST}$ between them. In general, if m:n mappings are used as input, a connected component could contain also more than two concepts. As mentioned in Section II, we will concentrate on 1:1 mappings, but the algorithm can be easily extended to m:n mappings introducing slight changes.

Let $L = \{l_1, ..., l_n\}$ be the set of the labels of the merging source concepts, we define the label of the integrated concept $C$ as $label(C) = genLabel(l_1, ..., l_n)$. The function $genLabel()$ can be defined in different ways but this is not the focus in this paper; for simplicity we define it as a string concatenation of the source labels and if two or more concepts have the same label, we consider their string value only once, adding a "*" symbol to mean that the label has more than one occurrence. For example, for a connected component with source labels: $L = \{Computer, PC, PC\}$, the generated label for the merged concept is $label = Computer\_PC^*$.

Another labeling function could return only one of the source labels (for example a target one if present) and add the remaining ones as a comment or description. A similar labeling system could be more helpful if one wants to reuse the merge result in an automatic matching task.

For two attributes involved in an attribute correspondence, we add only one attribute with a label depending on the labels of source attributes. We use a similar function to determine the label of merged attributes; e.g. for two corresponding attributes with label $Price$, we generate for the integrated concept only one attribute with label $Price^*$. In the running example, the integrated concept *Laptops HP* will have the following list of attributes $atts = \{Price^*, CPU, Display\}$. In Algorithm 2, the function *genAttributeList(CC)* generates the attribute list for an integrated concept starting from the set of source concepts in a connected component $CC$.

At this point, for each integrated concept $C$ we generate a new node in $I$.

Finally we translate the set of edges in $G_1$ and $G_2$ in a new set of labeled edges in $I$. In particular, for each edge $e_1 = C_1 \rightarrow C_2$ in $G_1$ (the source concept graph), we produce a *S-labeled edge* in $I$ – in the following called *S-edge* – defined as $s_e = C_{i1} \rightarrow C_{i2}$ where $C_{i1}$ and $C_{i2}$ are respectively the corresponding integrated concepts in $I$ of $C_1$ and $C_2$.

Similarly, for each edge $e_2 = D_1 \rightarrow D_2$ in $G_2$ (the target concept graph), we produce a *T-labeled edge* in $I$ – in the following called *T-edge* – defined as $t_e = D_{i1} \rightarrow D_{i2}$ where $D_{i1}$ and $D_{i2}$ are respectively the correspondent integrated concepts in $I$ of $D_1$ and $D_2$.

The integrated concept graph built for the example 1.1 is in Figure 3; $S_1$ and $T_1$ are respectively two examples of s-edge and t-edge.

It is easy to note that, given an integrated concept graph $I = (V, E)$ where $V$ is the set of nodes and $E$ is the set of edges, it is possible to split the set $E$ in two subsets, called $E(S)$ and $E(T)$, such that $E(S)$ contains all S-edges in $E$ and $E(T)$ contains all T-edges in $E$. Subsets $E(S)$ and $E(T)$ are a partition of the set $E$, since they are disjoint and their union is equal to $E$. This is also possible since we allow more than one edge between two nodes in $I$. Figure 4(a) shows an example where there exist two different edges, a s-edge and a t-edge, between the concepts $A^*$ and $B^*$.

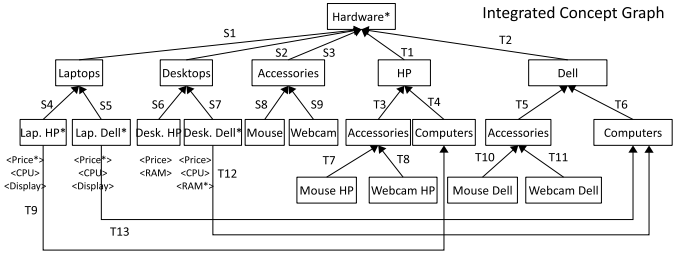

Fig. 3. Integrated Concept Graph

The *complexity* of the algorithm in the preliminary phase depends on the size and on the kind of source taxonomies. We assume an average number of concepts *n* and an average number of is-a relationships *r* per input taxonomy. If $O_1$ and $O_2$ are taxonomies with a simple hierarchy (i.e. with no multiple inheritance), the average number of relationships is $r = n - 1$. Analyzing the pseudo-code in Algorithms 1 and 2, the complexity of the preliminary phase is $O(r)$ and thus $O(n)$ if input taxonomies are hierarchies. If source taxonomies contain multiple inheritance, the number of edges can degenerate in the (highly unlikely) worst case to $\frac{n(n-1)}{2}$ resulting in complexity $O(n^2)$.

As mentioned, the preliminary phase of our algorithm is based on [5], but one of the main differences between these algorithms is the distinction in the integrated concept graph between edges coming from source and that ones coming from target. As we will discuss in the next phase of the algorithm, this distinction will be very important and helpful to visit the integrated concept graph and to automatically produce the final result.

### B. Main phase

The main phase is the most important step in the integrated ontology generation process. It is based on a graph visiting algorithm and the distinction between source and target edges in the integrated concept graph plays an important role.

Before describing the details of the algorithm, we need to introduce the notions of *source and target paths*.

First of all, we give the definition of *Path*, as known in graph theory:

**Definition [Path]** A path $P$ in a graph $G$ is a sequence of nodes such that, from each node there exists an edge between this node and the next one in the sequence. $P$ can be finite or infinite. A finite path has a *start node* and an *end node*; the other nodes are called *internal nodes*.

In this paper we consider only finite paths.

**Definition [Top Level Concepts]** An integrated concept $C$ is a *Source Top Level Concept* if $C$ has no outgoing s-edges in the graph $G$ but at least an incoming s-edge. This is equivalent to say that $C$ has one or more children but no parent with respect to s-edges. Similarly, $C$ is a *Target Top Level Concept* if it has no outgoing t-edges in $G$ but at least an incoming t-edge. Finally we define $C$ as a *Top Level Concept* if it is either a source or a target top level concept.

For example, in the integrated concept graph shown in Figure 3, the concept with label $Hardware^*$ is both a source and a target top level concept since it has only incoming s-edges and t-edges but no outgoing edges. As we will discuss later in this Section, $Hardware^*$ can be set as root of $I$ since there are no other top level concepts in $I$.

**Definition [TLC Path]** A finite path $P$ is called a *TLC path* if its end node is a Top Level Concept in the graph.

A path $P$ in a graph $G$ is a *cycle* if the start node and the end node are the same. Obviously the choice of start and end nodes is arbitrary. The integrated concept graph in Figure 4(b) has a cycle on the set of nodes $\{A^*, E, C^*, B\}$ and on the set of edges $\{S_1, S_2, T_1, T_2\}$.

Let $N$ be a node in a graph $G$ and $P$ a TLC path with start node $N$. $P$ is a *source-path* or simply *s-path* if it contains only s-edges. Similarly, $P$ is a *target-path* or simply *t-path* if it contains only t-edges. In Figure 3, for example, $P_1 = \{S_4 - S_1\}$ is a s-path, $P_2 = \{T_9 - T_4 - T_1\}$ is a t-path.

Now we are ready to discuss in detail the main phase of the algorithm. Given an integrated concept graph $I$, the algorithm generates an integrated taxonomy $T$. As for the preliminary phase, we identified several steps:

- Step 2.1: Removing cycles in $I$
- Step 2.2: Translation of t-edges
- Step 2.3: Translation of s-edges
- Step 2.4: Creating nesting structure

In the following we detail each step of the algorithm and we show the resulting intermediate and final results for our running example. The pseudo-code for the main phase is shown in Algorithm 3.

Step 2.1: [removing cycles; line 1 of Alg. 3] In this step we check if cycles are present in the graph $I$ in order to remove them. If we assume that the source and the target concept graphs $G_1$ and $G_2$ are cycle-free, any cycle in $I$ cannot involve only s-edges or t-edges. It is worth to note that there can be more than one way to solve this kind of cycles. For example, in [13] all concepts involved in a cycle are merged in a single concept since is-a relationships are transitive and a similar cycle implies equality of all its concepts. In our algorithm, for each cycle in $I$, we can break the cycle just deleting one of the s-edges involved in the cycle. The intuition behind this choice is that we define our algorithm as target-driven in order to preserve the target structure in the final result, and the removal of a s-edge does not modify the target structure. In this step, the user might choose which edge to remove for producing a better solution.

Figure 4(b) shows an example of a cycle in an integrated concept graph. In this example, the cycle can be broken in two different ways depending on which s-edge will be removed. The two possible results are shown in Figure 4(c): on the left side the solution obtained removing the s-edge $S_1$ and on the right side that one generated removing $S_2$. It is easy to note that the target structure is still maintained for both solutions even if the first solution seems to be better because it also maintains the position of the target top level concept. In order to generate the first solution instead of the second one, where possible, the algorithm automatically drops the s-edge outgoing from a target top level concept, such as $S_1$ in our example. If a top level concept is not involved in the cycle, a random s-edge will be removed. In the last case, a user could also interact with the system choosing which s-edge to delete in order to obtain a more specific result.

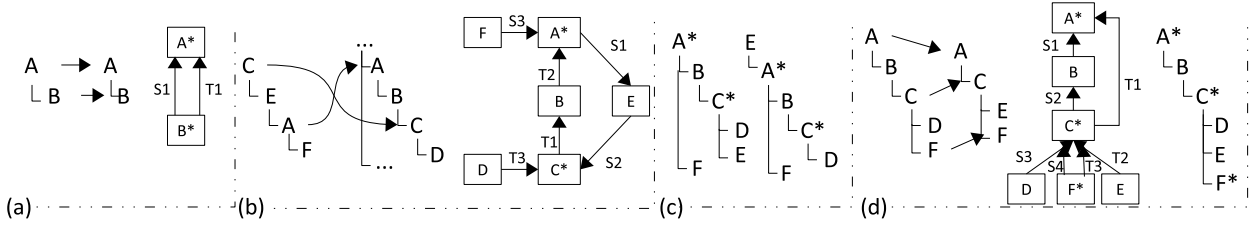

Fig. 4. (a) Two edges between the same concepts - (b) Example of a cycle - (c) Possible results after removing cycles - (d) An example of a more detailed source structure

**Algorithm 3** $IntegratedTaxGen(I)$
**Input:** an Integrated Concept Graph $I = (V, E)$
**Output:** an Integrated Taxonomy $T$
1: $RemoveCycles(I)$
2: **for each** t-edge $e = N_1 \to N_2$ in $I$ **do**
3:   **if** $\exists$ exactly one s-path $P$ from $N_1$ to $N_2$ s.t. $length(P) > 1$ **then**
4:     mark all edges in $P$ as *relevant*
5:   **else**
6:     create an is-a relationship between $N_1$ and $N_2$ in $T$
7:   **end if**
8: **end for**
9: $candidates \leftarrow \{X : X \in V \wedge X$ has at least one outgoing s-edge but no incoming s-edges$\}$
10: **for each** $X$ in $candidates$ **do**
11:   $spaths \leftarrow$ set of s-paths with start node $X$
12:   **for each** s-path $P$ in $spaths$ **do**
13:     **for each** edge $e = C_1 \to C_2$ in $P$ **do**
14:       **if** $C_1$ has no outgoing t-edges **then**
15:         mark $e$ as *relevant*
16:       **else**
17:         break
18:       **end if**
19:     **end for**
20:   **end for**
21: **end for**
22: **for each** relevant s-edge $e = N_1 \to N_2$ **do**
23:   create an is-a relationship between $N_1$ and $N_2$ in $T$
24: **end for**
25: $A \leftarrow$ set of all *relevant* source top level concepts in $I$
26: $B \leftarrow$ set of all target top level concepts in $I$
27: $TLCs \leftarrow A \cup B$
28: **if** $size(TLCs) = 1$ **then**
29:   $root(T) \leftarrow TLCs[0]$
30: **else**
31:   $root(T) \leftarrow$ create a new node $R$
32:   **for each** top level concept $tlc$ in $TLCs$ **do**
33:     nest $tlc$ in $root(T)$
34:   **end for**
35: **end if**
36: **return** $T$

Step 2.2: [translation of t-edges; lines 2-8 of Alg. 3] Recall that, given an integrated concept graph $I = (V, E)$, it is always possible to partition the set $E$ in two subsets $E(S)$ and $E(T)$ such that $E(S)$ contains all s-edges and $E(T)$ contains all t-edges in $E$. For each edge $e(t) = N_1 \to N_2$ in $E(T)$, we normally create an is-a relationship between the integrated concepts $C_1$ and $C_2$ corresponding to $N_1$ and $N_2$ in $I$, respectively, in order to maintain the target concepts and relationships. The only exception is when there exists exactly one source path $P$ with start node $N_1$ and end node $N_2$ containing more than one s-edge. In the latter case we do not create a direct relationship between $C_1$ and $C_2$ but we mark all edges in $P$ as relevant (for the merged taxonomy) so that they will be translated in the next step. The intuition behind this choice is that we want to preserve the target structure in the final result but if two concepts have a more detailed structure in the source, we want to reward it in the merged taxonomy since it preserves and extends the target structuring between $N_1$ and $N_2$, and this is possible thanks to the transitivity of is-a relationships. Let us consider, for example, the scenario shown in Figure 4(d). In the target taxonomy, concepts $A$ and $C$ are connected by a direct is-a relationship, while, in the source, they are connected through the concept $B$. It means that the source taxonomy has a higher level of detail than the target taxonomy and we want to report this in the final result, since the target relationship between $A$ and $C$ is still maintained, although indirectly. In fact, if $r(C, B)$ and $r(B, A)$ are two is-a relationships, the relationship $r(C, A)$ is implied.

Step 2.3: [translation of s-edges; lines 9-24 of Alg. 3] The translation of s-edges is the most important step in the algorithm, because the graph visit tries to integrate in $T$, in a "correct" position, the missing concepts coming from the source taxonomy. First of all, we define $candidates$ as the set of all nodes $L$ such that $L$ has at least one outgoing s-edge but no incoming s-edges. Informally speaking, we are looking for all leaf nodes in $I$ with respect to s-edges. For each node $L$ in $candidates$, we find all s-paths with start node $L$ and for each s-path $P$ we want to check which of its edges are relevant for the merge results without introducing redundancy in addition to the target edges that will be translated. We therefore traverse each s-path $P$ and consider its edges as relevant until one node in P has outgoing t-edge indicating that the remaining path is already covered by $T$. This criterion also observes the case when the remaining s-edges extend a t-edge since such edges would have been already identified as relevant in the previous step and we do not have to translate them again. In the example shown in Figure 4(d), the set of candidate nodes is *candidates* = $\{D, F^*\}$; the node $D$ has only one s-path $P = \{S_3, S_2, S_1\}$. The start node of $S_3$ - the first edge in $P$ - is $D$ and since it has no outgoing t-edges we mark $S_3$ as relevant; then we analyze $S_2$, its start node is $C^*$ and since $T_1$ is an outgoing t-edge for it, the algorithm stop because a translation for $C^*$ has already been proposed in the previous step. The same goes for the other candidate node $F^*$; in this case the algorithm stops at the first step since $F^*$ has an outgoing t-edge.

At this point we can decide which source nodes should be integrated in the final result and which not. We define a node $X$ as *not relevant* if all outgoing and incoming s-edges are marked as not relevant. For each s-edge $e = N_1 \to N_2$ marked as relevant and called $C_1$ and $C_2$ respectively the integrated concepts in $T$ corresponding to $N_1$ and $N_2$ in $I$, we create an is-a relationship between $C_1$ and $C_2$.

If we apply this step to our running example shown in Figure 3, the set of candidate nodes is *candidates* = {*Laptops HP\**, *Laptops Dell\**, *Desktops HP*, *Desktops Dell\**, *Mouse*, *Webcam*}. It is easy to note that edges $S_2$, $S_3$, $S_6$, $S_8$, $S_9$ will be marked as relevant and the remaining s-edges ($S_1$, $S_4$, $S_5$, $S_7$) as not relevant. The node *Laptops* is considered not relevant and it will not be translated in the integrated ontology. On the other side, an is-a relationship will be created, for example, between *Desktops* and *Desktops HP*.

Let $TLCs$ be the set of all Top Level Concepts. If $TLCs$ contains only one Top Level Concept, this will be the root of $T$, otherwise we create an artificial root node $R$ in $T$ and for each concept $C$ in $TLCs$ we create an is-a relationship from $C$ to $R$. Note that the set $TLCs$ contains only the relevant top level concepts; a source top level concept is defined as *relevant* if at least one incoming s-edge is relevant.

The integrated ontology $T$ generated by the algorithm for the Example 1.1 is shown in Figure 5. In order to keep the figure more compact, we omitted the attributes, since they were already reported in the integrated concept graph drawn in Figure 3. We observe that the target structure is fully preserved while the source taxonomy is only partially included since their concepts *Laptops HP*, *Laptops Dell* (and thus *Laptops*) as well as *Desktops Dell* are covered by corresponding target concepts.

The *complexity* of the main phase is also different if a single or a multiple inheritance is present in the sources. In the first case the complexity is still linear with respect to the sum of source concepts, in the latter case it is quadratic with respect to source concepts. The default result generated automatically by the system might not satisfy the subjectivity of the user. In this case, ATOM shows which concepts were considered as not relevant and the list of the target concepts with their paths that make them redundant in order to help the user to choose a different solution.

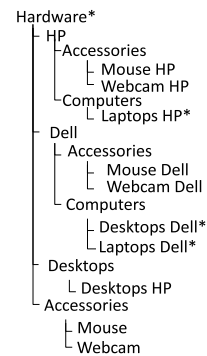

Fig. 5. Result of the base algorithm

### C. Multiple Inheritance

In order to show how this approach can be applied to ontologies with multiple inheritance, let us consider the example shown in Figure 6. It represents a small subset of the match scenario proposed by the Ontology Alignment Evaluation Initiative (OAEI) [1] that merges part of the subgraph describing *"Eye Muscles"* in the Mouse Anatomy with the subgraph describing a similar concept in the NCI Thesaurus. It is important to note that the graphs are much more complex in the original ontologies but, in order to keep this example smaller and more readable, we considered only some concepts. The merge scenario with the full ontologies will be discussed in Section VI.

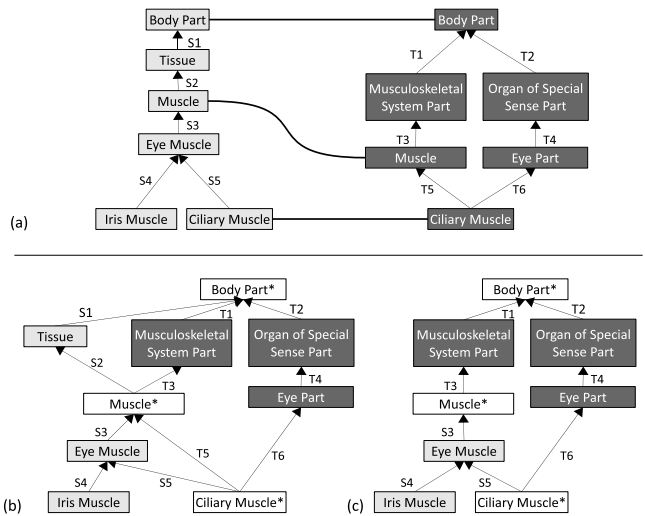

Fig. 6. Example with multiple inheritance: (a) the input taxonomies and equivalence mapping - (b) the integrated concept graph - (c) the merge result

The leaf concept *Ciliary Muscle* can be reached by different paths in the source and in the target; moreover it has multiple paths in the target ontology due to the multiple inheritance.

The integrated concept graph containing all the concepts and relationships coming from source and target taxonomies and the merge result for the given example is shown in Figure 6 (b) and (c). We highlighted with a white background the merge concepts and with a light and dark-gray color that concepts coming only from the source or from the target respectively, i.e. the concepts that are not covered by a correspondence in the input equivalence mapping.

For example, the source leaf concept *Iris Muscle* does not have an equivalent concept in the target but is still relevant for the merge result since it is a leaf in the source and could contain instances (see property (P3)). On the other hand, source concepts such as *Tissue* were not translated in the merge result otherwise a semantic overlap, defined in terms of multiple paths (see property (P4)), would be added for the concept *Ciliary Muscle*. In fact, in this case, the number of paths for this concept in the merge result is still the same than in the target ontology. Moreover, as we discussed in Section III, the algorithm rewarded the source path $\{S_5 - S_3\}$ since it preserves and extends the target structuring between the concepts *Ciliary Muscle* and *Muscle*. In fact the target is-a relationship defined by $T_5$ is implied by the is-a relationships in $S_5$ and $S_3$.

It is easy to see that the proposed algorithm generates a solution that meets the requirements (P1) to (P5) introduced in Section II for both examples. In particular, all the target concepts (P1) and is-a relationships (P2) are also in the merged taxonomy, all the source and the target leaf nodes were translated so that all instances are preserved (P3) and no semantic overlap was introduced in the result (P4), since no multiple paths were added for the leaf concepts and the tree-like structure was preserved for the example with single inheritance. The generation of mappings discussed in Section V shows that each leaf concept of the input taxonomies is mapped to the merged taxonomy and that if two concepts are equal in the equivalence mapping then they are mapped to the same merged concept in the result and vice versa (P5).

### D. Open Problems

By merging corresponding concepts we reduce semantic overlap compared to a simple union of the input taxonomies; furthermore we eliminate redundant inner nodes such as *Laptops* or *Tissue* for the running examples. Still there is remaining semantic overlap in the merge result determined by the base algorithm and we discuss next how to further reduce it with the extended algorithm.

Moreover, we assumed that in the ontology model only leaf nodes can contain instances and this requirement must be satisfied also for the merge result. If the input mapping contains an equivalence correspondence between a leaf node and an inner node, the corresponding merged concept in the result could be an inner node and some instances will be moved to it, which is not in accordance with the assumed model constraint. For this reason, a refinement step could be required in order to migrate these instances down the hierarchy to leaf nodes. Let us consider the example shown in Figure 7 representing two different product catalogs. In particular, the source taxonomy defines only one concept to classify the *Software* category while the target one defines different subcategories such as *Antivirus* or *Games*. As we will discuss in Section V, the generated equivalence mapping $Map_{ST'}$ between the source taxonomy and the merge result creates a correspondence $(Software, Software*)$ that would migrate the instances from the source leaf node *Software* to the merged inner node *Software\**. But this is not possible due to our assumption that instances are limited to the leaf nodes. In the next Section, we discuss how the specification of a more semantic mapping could solve these problems. For the basic algorithm we would have to extend the mappings $Map_{ST'}$ and $Map_{TT'}$ (see Section V) to move instances for concepts that are no leaf concepts in $T'$ to the leaf level, e.g. to migrate *Software* instances of $T$ to the leaf concepts *Antivirus* and *Games*.

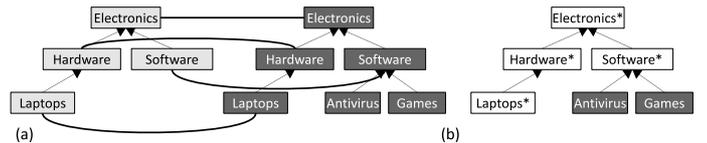

Fig. 7. Example with equivalence correspondences between a leaf and an inner node

## IV. EXTENSIONS

Inspecting the result of the base algorithm shown in Figure 5, reveals that not all concepts seem well placed and that there is still some semantic overlap due to the differences in the original taxonomies. For example the concept *Desktops HP* should not be in a different subtree than concept *HP*. Furthermore, there is likely overlap between the general *Mouse* concept under *Hardware\*. Accessories* and the more specific concepts *Mouse HP* and *Mouse Dell* (similarly for *Webcams*). The base algorithm could not better deal with such cases since the semantic relationships between these concepts have not been expressed in the provided equivalence mapping (see Fig.1). Hence a prerequisite to improve the merge result is the provision of more semantic mappings between the input taxonomies, in particular is-a and inverse-isa relationships in addition to equivalence relationships. For the running example, we can then specify that *Desktop HP* in the source taxonomy "is-a" *HP Computer* in the second taxonomy and that *Mouse* in the source represents every kind of mouse and not a mouse of a specific brand, as instead in the target is. So we could say that *Mouse* in the source is a superclass for both *Mouse HP* and *Mouse Dell* in the target.

The extended merge algorithm that we present in this section is based on such enriched input mappings consisting of equivalence, is-a and inverse-isa relationships.

Figure 8 shows an *is-a* and an *inverse-isa* mappings for the running example introduced in Section I. In order to keep the figure more readable, we omitted the equivalence mapping but it is still a necessary input for the algorithm. An is-a mapping between two ontologies is defined as a set of *is-a correspondences*. An is-a correspondence is an oriented correspondence from a source concept to a target concept. Similarly we define an *inverse-isa mapping* as a set of *inverse-isa correspondences* from source to target concepts.

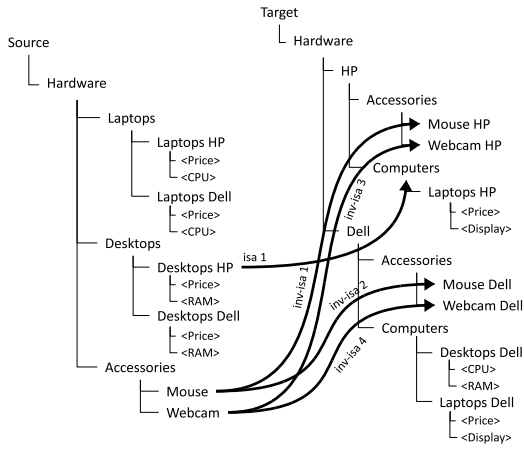

Fig. 8. Example of *is-a* and *inverse-isa* mappings

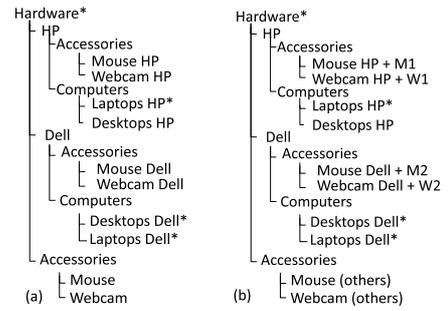

Fig. 10. (a) Result of isa-mapping translation - (b) Result of inverse-isa mapping translation

Semantically we could define an "inverse-isa" relationship as the inverse of an "is-a" relationship, e.g. the relationship *Mouse* inverse-isa *Mouse HP* is semantically equivalent to the relationship *Mouse HP* is-a *Mouse*. However, since our algorithm is target-driven we need different approaches for dealing with is-a and inverse-isa relationships so that we want to emphasize the distinction.

The preliminary phase of the extended algorithm is basically the same than for the base algorithm. The difference is that the integrated concept graph, output of this phase, contains not only s-edges and t-edges but also two new sets of edges: *isa-edges* and *inv-isa-edges* representing respectively is-a and inverse-isa correspondences in the auxiliary mapping. Figure 9 shows the integrated concept graph for the Example 1.1; isa-edges and inv-isa-edges are represented by curved lines.

Because of the semantically different nature of an "is-a" and an "inverse-isa" mapping, they will be translated in two different ways. In the following we describe how we modified the base algorithm to manage is-a and inverse-isa mappings.

*1) Translation of is-a mapping:* The intuition behind the translation of an isa-edge is that it represents a subclass relationship between a source concept and a target concept and we want this relationship to be translated also in the integrated ontology.

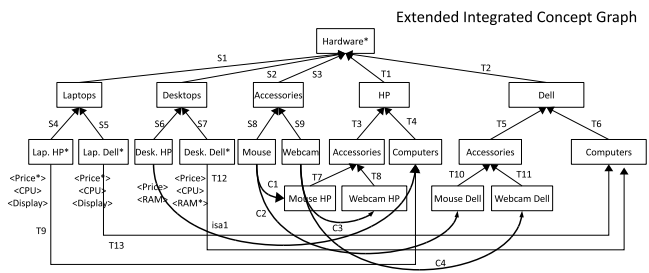

Fig. 9. Extended Integrated Concept Graph with is-a and inverse-isa mappings

Before describing how we changed the base algorithm, it is necessary to introduce some new definitions.

**Definition [isa-path]** Let $N$ be a node in a graph $I$ and $P$ a TLC path with start node $N$. $P$ is an *isa-path* if it contains only isa-edges. Let $N$ be a node in a graph $G$ and $P$ a TLC path with start node $N$. $P$ is a *mixed-path* if it contains only t-edges or isa-edges. Let $N$ be a node in a graph $G$ and $P$ a path with start node $N$. $P$ is a *combined-path* or *c-path* if it contains only s-edges or isa-edges.

In order to translate isa-edges in the final result, we introduce a new step in the main phase after the translation of the target edges.

Step 2.2.1: [translation of isa-edges; lines 9-11 of Alg. 4] For each isa-edge $e = N_1 \to N_2$, called $C_1$ and $C_2$ respectively the integrated concepts in $T$ corresponding to $N_1$ and $N_2$ in $I$, we create an is-a relationship between $C_1$ and $C_2$. If we consider the integrated concept graph shown in Figure 9, at the end of this step, we translate the single isa-edge with label "$isa_1$" nesting the concept *Desktops HP* in *Computers*.

It is easy to note that the presence of isa-edges in the integrated concept graph influences the translation of source edges, since a source concept could have been already translated if involved in a t-path but also in an isa-path (or a mixed-path). For this reason, we changed the translation of s-edges accordingly as follows:

Step 2.3: [translation of s-edges; lines 12-27 of Alg. 4] This step is similar to that one defined in the base algorithm with the difference that for each source leaf node we have to check if there exist other paths containing at least one t-edge or one isa-edge and to mark s-edges as relevant or not relevant accordingly. In the example shown in Figure 9, the set of candidate nodes is *candidates* = {*Laptops HP*\*, *Laptops Dell*\*, *Desktops HP*, *Desktops Dell*\*, *Mouse*, *Webcam*}. Since only the s-paths with start nodes *Mouse* and *Webcam* have no mixed-paths, we mark the edges $S_3$, $S_8$, $S_9$ as relevant and the remaining s-edges ($S_1$, $S_2$, $S_4$, $S_5$, $S_6$, $S_7$) as not relevant.

The algorithm remains basically unchanged in the next steps, but the final result is evidently different and it is shown in Figure 10(a). As we can see, the concept *Desktops HP* is now correctly placed; the source concept *Desktops* is no more relevant and it was dropped from the final result.

**Algorithm 4** $ExtendedIntegratedTaxGen(I)$

**Input:** an Integrated Concept Graph $I = (V, E)$
**Output:** an Integrated Taxonomy $T$
1: $RemoveCycles(I)$
2: **for each** t-edge $e = N_1 \rightarrow N_2$ in $I$ **do**
3:    **if** $\exists$ exactly one c-path $P$ from $N_1$ to $N_2$ s.t. $length(P) > 1$ **then**
4:      mark all edges in $P$ as *relevant*
5:    **else**
6:      create an is-a relationship between $N_1$ and $N_2$ in $T$
7:    **end if**
8: **end for**
9: **for each** isa-edge $e = N_1 \leftarrow N_2$ in $I$ **do**
10:    create an is-a relationship between $N_1$ and $N_2$ in $T$
11: **end for**
12: $candidates \leftarrow \{X : X \in V \land X$ has at least one outgoing s-edge but no incoming s-edges$\}$
13: **for each** $X$ in $candidates$ **do**
14:    $spaths \leftarrow$ set of s-paths with start node $X$
15:    **for each** s-path $P$ in $spaths$ **do**
16:      **for each** edge $e = C_1 \rightarrow C_2$ in $P$ **do**
17:        **if** $C_1$ has no outgoing t-edges or isa-edges **then**
18:          mark $e$ as *relevant*
19:        **else**
20:          break
21:        **end if**
22:      **end for**
23:    **end for**
24: **end for**
25: **for each** relevant edge $e = N_1 \leftarrow N_2$ **do**
26:    create an is-a relationship between $N_1$ and $N_2$ in $T$
27: **end for**
28: $A \leftarrow$ set of all *relevant* source top level concepts in $I$
29: $B \leftarrow$ set of all target top level concepts in $I$
30: $TLCs \leftarrow A \cup B$
31: **if** $size(TLCs) = 1$ **then**
32:    $root(T) \leftarrow TLCs[0]$
33: **else**
34:    $root(T) \leftarrow$ create a new node $R$
35:    **for each** top level concept $tlc$ in $TLCs$ **do**
36:      nest $tlc$ in $root(T)$
37:    **end for**
38: **end if**
39: **for each** inv-isa-edge $e = N_1 \rightarrow N_2$ in $I$ **do**
40:    **if** exists a relevant s-edge or t-edge $r = N_2 \rightarrow N_1$ **then**
41:      create a new concept $N_{others}$ with $label(N_{others}) \leftarrow label(N_1) + "(others)"$ (if does not exist a similar concept)
42:      create a new is-a relationship between $N_{others}$ and $N_1$
43:    **else**
44:      **if** $N_1$ has not yet been renamed **then**
45:        $label(N_1) \leftarrow label(N_1) + "(others)"$
46:      **end if**
47:      $label(N_2) \leftarrow label(N_2) + subset(N_1)$
48:    **end if**
49: **end for**
50: **return** $T$

*2) Translation of inverse-isa mapping:* In order to manage inverse-isa mapping, we introduce a final step in the main phase algorithm where we rename labels of the concepts involved in inverse-isa-correspondences.

An inverse-isa-mapping semantically describes how a source concept can be split in two or more target concepts and how its instances should be partitioned. For example, in Figure 8 the source concept *Mouse* representing all kinds of mouse without distinction of brand, can be split in two target concepts *Mouse HP* and *Mouse Dell*. But, in general, *Mouse* can also contain mice with a brand that is different both from HP and Dell (e.g. Logitech Mouse) and we must be careful to not miss this information. We add the following step to the algorithm as a final step:

*Step 2.5: [translation of inv-isa edges; lines 39-49 in Alg. 4]* For each inv-isa-edge defined as $e = A \rightarrow B$, called respectively $label(A)$ and $label(B)$ the labels of nodes $A$ and $B$, we distinguish two cases: a) if there exists a s-edge or a t-edge between the same nodes but in opposite direction, a new concept with label equals to $label(A) + "(others)"$ will be added to the result and a new is-a relationship between this concept and $A$ will be created; b) if such an edge does not exist, we simply rename $label(A)$ in $label(A) + "(others)"$ and $label(B)$ in $label(B) + subset(A)$. The "+" operator indicates here a simple concatenation function; the $subset()$ function indicates a subset of the original set when applied some special conditions. For example, supposing that the *Mouse* concept has an attribute called *brand*, we could apply a filter condition based on the value of the *brand* attribute. We will discuss this point in detail in Section V.

Let us consider the two inverse-isa-correspondences $c_1$ and $c_2$ in our example:

$$c_1 = Mouse \rightarrow Mouse\ HP$$
$$c_2 = Mouse \rightarrow Mouse\ Dell$$

The renaming step will produce the following result:

| *Mouse* | $\Rightarrow$ | *Mouse (others)* |
| *Mouse HP* | $\Rightarrow$ | *Mouse HP* $+ subset_{HP}(Mouse)$ |
| *Mouse Dell* | $\Rightarrow$ | *Mouse Dell* $+ subset_{Dell}(Mouse)$ |

It is easy to note that the concept *Mouse (others)* represents the following set:

$$Mouse(others) \equiv M \setminus (subset_{HP}(M) \cup subset_{Dell}(M))$$

where $M$ is the original concept *Mouse*.

By analyzing source instances it is possible to discover if the concept *Mouse(others)* is empty or not. In particular, if $\{subset_{HP}, subset_{Dell}\}$ is a partition of the original concept *Mouse*, the node *Mouse(others)* is empty and could be removed from the merged taxonomy. But in general, if we did not consider this node, we could lose some information.
A similar result is obtained for the inverse-isa-correspondences $c_3$ and $c_4$ related to the *Webcam* concept.

Figure 10(b) shows the integrated ontology produced by the extended algorithm, after the renaming step.

The extended merge algorithm presented in this Section can solve also the problem about the migration of instances in merged inner nodes that we addressed in Section III-D. In the example shown in Figure 7, the user might refine the input mapping defining two inverse-isa correspondences $(Software, Antivirus)$ and $(Software, Games)$. In such a similar case, the source instances from the leaf node *Software* would be split and migrate to the merge leaf nodes *Antivirus* and *Games*. Moreover, a new concept with label *Software (others)* would be created as a child of the merged concept *Software\**. In Section V we discuss how the equivalence mappings between input taxonomies and the merged taxonomy can be automatically generated and in particular how it is possible to split instances involved in inverse-isa relationships defining specific filter conditions.

The full pseudo-code for the extended algorithm is shown in Algorithm 4. It is easy to note that steps introduced in the extended algorithm do not increase the complexity of the base algorithm.

## V. MAPPING GENERATION

In this section we discuss how to automatically generate equivalence mappings between the input taxonomies and the merged taxonomy as determined by the extended algorithm. The process is fully automatic and based on the extended integrated concept graph reflecting the equivalence, isa and inverse-isa relationships between the input taxonomies. These relationships produce different edges in the integrated concept graph and consequently different concepts and relationships in the merged taxonomy. In particular, correspondences in an equivalence mapping describe how two or more source concepts should be merged in the integrated taxonomy; on the other side, an isa-mapping does not produce merged concepts in the result, but it defines a subclass relationship between a source and a target concept, describing which should be the father of a source concept in the merged taxonomy; finally, an inverse-isa-mapping describes how to split a source concept - and its instances - in two or more concepts in the final result. Algorithm 5 shows how mappings $M_1$ and $M_2$ for relating the input ontologies $O_1$ and $O_2$ to the integrated ontology, respectively, are determined. The algorithm will determine in $M_1$ a correspondence for every target concept (since the merge algorithm is target-maintaining) and in $M_2$ a correspondence for every source concept explicitly reflected in the merged taxonomy. In particular there will be a correspondence for every leaf concept specifying where instances should migrate in the merged taxonomy.

Input equivalence correspondences and isa-correspondences can be translated at the same time analyzing nodes marked as relevant in the integrated concept graph $I$. As defined in Section III (Step 1.3), each integrated concept in $I$ contains the collections of source and target concepts from which it was generated. In this way, given an integrated concept $C$, it is always possible to know if $C$ was present only in the source, only in the target or in both. In the following, we call respectively $sc_c$ and $tc_c$ the collections of source and target concepts for an integrated concept $C$. It is important to note that we assume that all integrated concepts with a nonempty $tc_c$ are marked as relevant by default - i.e. all concepts present in the target taxonomy are relevant and must be translated in the merged taxonomy.

As described in Algorithm 5, for each relevant integrated concept $C_I$ in $I$ such that $C$ has no outgoing inv-isa-edges (they will be translated later in a different way), called $sc_c$ and $tc_c$ the sets of source and target concepts, for each concept $C$ in $sc_c$, we create a correspondence between $C$ and $C_I$; the same goes for concepts in $tc_c$.

For example, if we consider the integrated concept *Laptops HP\** in Figure 9, the sets $sc_c$ and $tc_c$ are respectively $sc_{Laptops\_HP^*}$ = {*Laptops HP*} and $tc_{Laptops\_HP^*}$ = {*Laptops HP*} and we will create a correspondence between the source concept *Laptops HP* and the integrated concept *Laptops HP\** and another one between the target concept *Laptops HP* and *Laptops HP\**. If we consider, instead, the node *Desktop HP*, $sc_{Desktops\_HP}$ = {*Desktops HP*} while $tc_{Desktops\_HP}$ is empty and only one correspondence between the source concept *Desktops HP* and the integrated concept *Desktops HP* will be generated.

The translation of inverse-isa-correspondences is a delicate step in the mapping generation process since each correspondence describes how to move only "some" instances of a source concept and not all instances related to it. In our running example, as discussed in Subsection IV-.2, correspondences with label $c_1$ and $c_2$ state that instances in the source concept *Mouse* should be split in three disjoint subsets: *Mouse HP*, *Mouse Dell* and *Mouse (others)*. As proposed in [10], a correspondence can have an attached filter - called *filter condition* - that states under which conditions the correspondence must be applied. For example, supposing the *Mouse* concept has an attribute called *brand*, we can define the following filter conditions respectively for correspondences $c_1$ and $c_2$:

$$fc_1 = [Mouse.brand='HP']$$
$$fc_2 = [Mouse.brand='Dell']$$

Filter conditions can be automatically generated by a matching tool when correspondences are created or they can be manually defined by a user; as discussed for correspondences, we do not investigate how filter conditions are generated since they are part of our algorithm input. At this point, we are ready to present how inverse-isa-correspondences are translated in the mapping generation process. As shown in Algorithm 5, for each node $C_I$ with at least one outgoing inv-isa-edge, called $C$ the corresponding source concept and called *c-edges* the list of all outgoing inv-isa-edges for $C_I$, we create a correspondence for each edge $e$ in *c-edges* and we attach the filter condition $fc$ defined on $e$. Finally, we create a correspondence between $C$ and the integrated concept with label "*(others)*" attaching a filter condition $fc_{others}$ defined as the negation of all filter conditions attached on inv-isa-edges outgoing from $C_I$.

If we consider again the concept *Mouse* in our running example, the set of outgoing inv-isa-edges is {$c_1$, $c_2$} with filter conditions $fc_1$ and $fc_2$ defined above. With respect to $c_1$ edge, we create a correspondence between the source concept *Mouse* and the integrated concept *Mouse HP* attaching $fc_1$ on it; similarly for $c_2$. Finally, we create a correspondence between *Mouse* and *Mouse (others)* and we define a new filter

**Algorithm 5** $MappingGen(I, T)$

**Input:** an Int. Concept Graph $I$ and a merged taxonomy $T$
**Output:** two equivalence mappings $M_s$ and $M_t$
1: $M_s \leftarrow$ empty
2: $M_t \leftarrow$ empty
3: $E_0 \leftarrow$ all merged nodes in $I$
4: $E_1 \leftarrow$ relevant nodes in $I$ with no outgoing inv-isa-edges
5: $E_2 \leftarrow$ nodes in $I$ with at least one outgoing inv-isa-edge
6: **for each** node $C_I$ in $E_0 \cup E_1$ **do**
7:    $sc \leftarrow$ set of source concepts for $C_I$
8:    $tc \leftarrow$ set of target concepts for $C_I$
9:    **for each** concept $C$ in $tc$ **do**
10:      create a correspondence $C - C_I$ in $M_t$
11:    **end for**
12:    **for each** concept $C$ in $sc$ **do**
13:      create a correspondence $C - C_I$ in $M_s$
14:    **end for**
15: **end for**
16: **for each** node $C_I$ in $E_2$ **do**
17:    $fcs \leftarrow$ empty
18:    $c\_edges \leftarrow$ set of all inv-isa-edges outgoing from $C_I$
19:    **for each** edge $e = C_1 \rightarrow C_2$ in $c\_edges$ **do**
20:      create a corr. $C_2 - C_I$ in $M_s$ with a filter cond. $fc_e$
21:      add $fc_e$ to $fcs$
22:    **end for**
23:    create a corr. $C_1 - C_1(others)$ in $M_s$ with a filter cond. "not $fcs$"
24: **end for**
25: **return** $M_s, M_t$

condition as follow:

$$fc_{others} = [Mouse.brand \neq \text{'HP'} \text{ AND } Mouse.brand \neq \text{'Dell'}]$$

The last correspondence states that all mice with a brand different from *HP* and *Dell* must be moved to the concept *Mouse (others)* in the merged taxonomy.
It is important to note that if concepts in the input taxonomies have attributes, we create an attribute correspondence between each of them and the corresponding attribute in the integrated concept. We omit details of this process here.
Figure 11 and 12 show the mappings $M_1$ and $M_2$ generated for our running examples; we have drawn correspondences between leaf nodes with a solid line and that ones between inner nodes with a dotted line. It is easy to see that properties (P3), (P4) and (P5) introduced in Section II are satisfied with respect to instances, since there is a correspondence for each leaf node in source taxonomies and each instance migrates to exactly one concept in the merged taxonomy; moreover if there exists an equivalence correspondence between two concepts in the input mapping, they are mapped to the same merged concept.

## VI. EVALUATION

Evaluating the result of an ontology merging algorithm is a complex task since there exist no ideal results. Often the quality of a merged taxonomy is subjective and it depends on the reference context and on the application domain. Moreover no benchmarks are available to evaluate a merging algorithm as, on the contrary, it is for schema mapping systems [2] or matching tools [1]. Since there are no ideal merge taxonomies with which to compare the result of our algorithm, it is not possible to evaluate it using standard quality measures, like Precision, Recall or F-Measure.

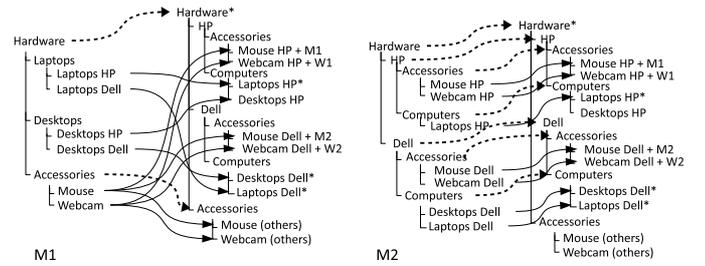

Fig. 11. Mapping $M_1$ and $M_2$

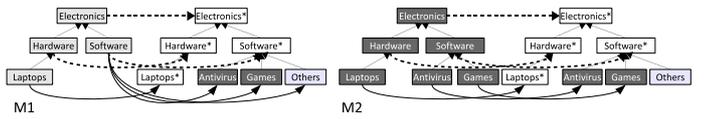

Fig. 12. Mapping $M_1$ and $M_2$

In this section we evaluate our algorithm and we study its performance on scenarios of various kinds and sizes.

The algorithms proposed in the paper have been implemented in the ATOM system [16], a working prototype written in Java offering a GUI to explore all steps of the merging generation process. ATOM has been integrated with COMA++ [3] which permits semi-automatic generation of input mappings. In this section we study the performance of our algorithm on scenarios of various kinds and sizes. We show that algorithms proposed efficiently compute a merged taxonomy even for large real-life ontologies. All experiments have been executed on an Intel Xeon machine with 2.66Ghz processors, 4 GB of RAM and a 64 bit operating system. We have used the prototype to run a number of experiments both on synthetic and real-life scenarios. Table I summarizes the list of experiments. We expressed the size of taxonomies as their number of concepts and the dimension of the integrated concept graph as the number of nodes and edges. Synthetic ontologies, named as *S1, S2 and S3*, have been manually created and designed to check the correctness of the algorithm and for this reason they have a small number of concepts and relationships (in Table I we reported only the most relevant ones); on the other side, real-life ontologies were chosen as very large ontologies and contain, on average, from hundreds to over 20000 concepts and they come from different domains. In particular, *Dmoz-Google Freizeit* and *Google-Web Lebensmittel* merge a part of DMoz and Google web directories; the *Anatomy (Mouse-NCI)* scenario merges the AdultMouseAnatomy (over 2700 concepts) with the anatomical part of the NCI Thesaurus (NCIT) (about 3300 concepts) and finally we considered different versions of eBay product catalog and we run the algorithm to merge them and find out possible differences. These versions of the catalog contain in average more than

22000 concepts and full mappings contain more than 20000 correspondences.

| Experiment Name | Source size | Target size | # of correspondences | Int.Concept Graph Size (nodes/edges) | Output Dimension (# concepts) |
|---|---|---|---|---|---|
| S1 | 9 | 11 | 6 | 14/19 | 13 |
| S2 | 12 | 15 | 12 | 15/20 | 15 |
| S3 | 15 | 10 | 10 | 15/18 | 11 |
| Dmoz-Google Freizeit | 72 | 68 | 68 | 72/138 | 72 |
| Google-Web Lebensmittel | 58 | 52 | 32 | 78/105 | 70 |
| Anatomy (Mouse-NCI) | 2744 | 3304 | 988 | 5060/6611 | 4996 |
| Ebay-v92-v93 | 24480 | 22513 | 23805 | 23188/44946 | 23123 |
| Ebay-v93-v94 | 22513 | 21034 | 21750 | 21797/41917 | 21713 |

TABLE I
SUMMARY OF EXPERIMENTS

*S1* shows a scenario with a partial overlap between sources - only six target concepts are mapped. The algorithm generated a merged taxonomy with 13 concepts starting from an integrated concept graph with 14 nodes; it means that considering the 9 source concepts, 6 of them have been merged, 2 have been translated in the result and only one was recognized as not relevant and dropped. Scenario *S2* is an example of full overlap - the source taxonomy is fully contained in the target one - and the result is equivalent to the target taxonomy. Finally, the experiment *S3* shows a scenario where the target is fully contained in the source one; in this case the result contains all target concepts and only one coming from the source taxonomy while the remaining ones were marked as not relevant.

About the real-life scenarios, we distinguished three categories: web directories, life science and product catalogs. The web directory category contains two experiments of different nature. The first one presents a full overlap between input taxonomies; all the concepts not mapped in the source are marked as relevant and translated in the final result. The second scenario presents a partial overlap and in this case the algorithm generated a result containing about 60% of not mapped source concepts.

The scenario in life science category (*Mouse-NCI*) merges two medium-scale taxonomies with 3000 concepts in average. It is possible to note how the algorithm considered some source concepts as not relevant in the final result. Mouse and NCI ontologies contain both is-a and part-of relationships but, in our experiments, we considered only is-a relationships.

Finally, scenarios in product catalog category address some issues. If we consider a mapping containing only correspondences between products moved in a different category, the dimension of the merged taxonomy is exactly the same of the integrated concept graph and it means that all concepts are relevant. On the other side, if we consider a full mapping between input taxonomies as reported in Table I, the merge result is much more compact since more concepts are merged and some source concepts are marked as not relevant.

Now we discuss the scalability of the algorithm on ontologies of large size. As already discussed in previous sections, the complexity of the algorithm is theoretically linear with respect to the number of concepts for taxonomies with single inheritance and quadratic in presence of multiple inheritance. We experimentally proved that the algorithm has good performance also with real and large taxonomies. In particular, ontologies in the anatomy scenario have multiple inheritance, while each concept in eBay product catalogs has only one parent. We measured execution time for each step in the algorithm, in particular the time necessary to generate concept graphs, matching graph, integrated concept graph and finally that one necessary to visit the integrated graph and produce the final result. We also measured the execution time to generate the final equivalence mappings between input and merge result. We reported in Fig. 13 execution times for medium and large-size scenarios and we omitted the other ones since the total execution time was much lower. We also omitted the concept graphs and the matching graph generation times, since they resulted not relevant with respect to the total time. The system generated the final result in about one second for the medium-size scenario with multiple inheritance (Mouse-NCI), and in less than 10 seconds for the large-size ones (eBay product catalogs), showing a very good scalability of the algorithms.

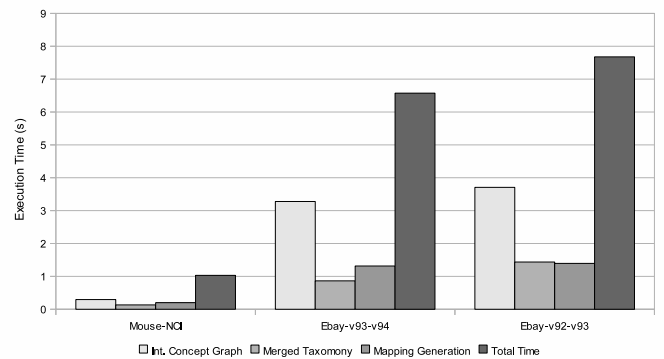

Fig. 13. Execution times on large-scale scenarios

## VII. RELATED WORK

In this section we review some related works in the fields of schema and ontology merging.

Several approaches have been proposed to merge schemas based on a pre-determined match mapping [13], [5], [17], [12], [18], [14] but, at the best of our knowledge, our approach is the first one to use more semantic mappings, like isa or inverse-isa mappings, to improve the merge result. Moreover, in contrast to previous approaches we do not try to preserve all non-matching concepts and their relationships in both input ontologies since this could introduce a semantic overlap in the merge result due to a different and overlapping structuring of the domain of interest.

In [13] the problem of merging two models given a set of correspondences is investigated. They study the problem from a more general point of view since their approach can be applied to a generic model such as a database schema, a UML model or an ontology defining a set of properties, called Generic Merge Requirements, that a merge model should satisfy. In this paper we reuse some of these requirements adapting them to our target-driven algorithm and define new properties.

[5] introduces an interactive algorithm to generate integrated schemas. Given a set of two or more source schemas - XML or relational schemas - and an equivalence mapping, it generates and enumerates multiple integrated schemas and requires user intervention to refine the result. Our approach is similar to that one proposed in [5] since we use a similar model based on concept and matching graphs but with significant differences. One of the main differences is the distinction between source and target edges that we introduced in the integrated graph since it plays an important role to generate the merge result in a fully automatic manner. For each integrated schema returned to the user, they also generate a mapping from the source schemas to the integrated schemas.

The approach proposed in [12] takes as input a set of relational schema and an equivalence mapping expressed under the form of conjunctive queries. As in [5], the proposed algorithm generates a mapping between input schemas and mediated schema. One of the main goals is to provide a set of features that the merge result and generated mapping should have.

A more automatic approach to schema integration is proposed in [14]. The approach is based on the use of directed and weighted correspondences and they use this information to rank the integrated schemas generated by the algorithm. It also provides an interactive step where a user can add one or more constraints in order to refine the final result.

A different approach was studied in [18], based on the notion of a probabilistic mediated schema, i.e. a set of integrated schemas with a related probability; they propose an automatic pay-as-you-go approach where the system proposes a starting mapping that can be incrementally improved if necessary.

Finally, the system presented in [17], addresses both the matching and the merging problem for XML schemas. They propose an incremental approach that discovers equivalence correspondences between the sources and, at the same time, incrementally generates a merged schema and the mapping between sources and mediated schema.

Previous approaches on ontology merging [11], [9], [19] have been primarily focused on the problem of ontology alignment, or ontology matching, and do not use the separation of matching and merging.

PROMPT[11] proposes an algorithm for aligning and merging ontologies. The merging algorithm is semi-automatic since it can perform some tasks automatically, while the other tasks are suggested to the user as a list of possible operations to execute. The user intervention can generate inconsistencies in the merged ontology; the system can discover these inconsistencies and suggest to the user a possible solution.

Chimaera[9] is one of the tools implemented in the Ontolingua system [7], a tool to design and manipulate ontologies. Chimaera guides the user in the merging process proposing a list of possible operations to perform, like in [11].

FCA-Merge[19] presents a merging algorithm based on a lattice of concepts. The lattice is automatically derived but the generation of the merged ontology requires the user intervention to explore the lattice.

The problem of merging multiple taxonomies based an a given alignment was addressed in [20]. In particular they investigate the merging problem when the relationships between concepts of different taxonomies are expressed as RCC-5 constraints. Unlike the other similar systems above, it generates the merge result in one step.

Finally, an evaluation of ontology merging tools was presented in [8]. They compared PROMPT and Chimaera tools on merging bio-ontologies. However, the focus was not on the merge result but on the system itself and the user effort to obtain a specific result.

A different approach to the evaluation of the quality of an integrated schema is discussed in [6]. Given an ideal schema built by an expert user, they propose a new metric called *Schema proximity* of two schemas that measures the quality of the generated schema in terms of *Structurality*, *Completeness* and *Minimality*. But as we discussed in Section VI, the quality of a merged taxonomy is subjective and more than one ideal solution could exist; moreover, it is hardly possible to manually generate an ideal merged schema for very large scenarios.

## VIII. CONCLUSIONS

We proposed a base algorithm and an extended algorithm for automatically merging a source taxonomy into a target taxonomy. The base algorithm utilizes equivalence mappings between the input taxonomies, while the extended approach additionally uses is-a and inverse-isa relationships. Our target-driven approach preserves all concepts and relationships of the target taxonomy and can largely limit the semantic overlap in the merged taxonomy especially for the extended algorithm. Furthermore, we determine mappings between the input taxonomies and the merge result that can be used for instance migration. The proposed algorithms have linear complexity for hierarchical taxonomies. The algorithms have been implemented in the ATOM system and could be successfully applied to large real-life taxonomies from different domains. ATOM generates a default solution in a fully automatic way that may interactively be adapted by users if needed.

In future work we will generalize our approach to support m:n mappings and other kinds of relationships in addition to is-a relationships; moreover we will extend our algorithms to support instances not only for leaf concepts but also for inner concepts. Furthermore, we plan to develop a benchmark for evaluating ontology merging algorithms which we consider as an open challenge.